\documentclass[12pt]{article}
\usepackage{amsthm,latexsym,amsmath,amssymb,amsfonts,color,colordvi,bbm}
\usepackage[mathscr]{eucal}
\usepackage{hyperref}
\numberwithin{equation}{section}

\definecolor{DarkViolet} {rgb}  {0.580392,0.000000,0.827450}
\definecolor{ForestGreen}{rgb}  {0.100000,0.408823,0.100000}
\definecolor{green3}     {rgb}  {0.000000,0.803921,0.000000}
\definecolor{mydarkblue} {rgb}  {0.282352,0.239215,0.803921}
\definecolor{OrangeRed}  {rgb}  {1.000000,0.270588,0.000000}
\definecolor{red3}       {rgb}  {0.803921,0.000000,0.000000}
\begin{document}
%%%%%%%%%%%%%%%%%%%%%%%%%%%%%%%%%%%%%%%%%%%%%%%%%%%%%%%%%%%%%%%%%%%%%%%%
    %             \version\versionno
    %             \draft
                  % \thispagestyle{empty}  %% remove for journal
\begin{flushright}
{\sf ZMP-HH/09-21}\\
   {\sf Hamburger$\;$Beitr\"age$\;$zur$\;$Mathematik$\;$Nr.$\;$349}\\[2mm]
   October 2009
\end{flushright}
\vskip 2.3em
\begin{center}\Large
TWENTY-FIVE YEARS OF TWO-DIMENSIONAL \\[2pt]  RATIONAL CONFORMAL FIELD THEORY
\end{center}\vskip 2.1em
\begin{center}
  ~J\"urgen Fuchs\,$^{\,a}$,~
  ~Ingo Runkel\,$^{\,b}$~ ~and~
  ~Christoph Schweigert\,$^{\,b}$~
\end{center}

\vskip 5mm

\begin{center}\it$^a$
  Teoretisk fysik, \ Karlstads Universitet\\
  Universitetsgatan 21, \ S\,--\,651\,88\, Karlstad
\end{center}
\begin{center}\it$^b$
  Organisationseinheit Mathematik, \ Universit\"at Hamburg\\
  Bereich Algebra und Zahlentheorie\\
  Bundesstra\ss e 55, \ D\,--\,20\,146\, Hamburg
\end{center}
\vskip 3em
 %% \noindent{\sc Abstract} \\[3pt]
 \centerline{\em
A review for the 50th anniversary of the Journal of Mathematical Physics. 
}

\vskip 2.8em

%%%%%%%%%%%%%%%%%%%%%%%%%%%%%%%%%%%%%%%%%%%%%%%%%%%%%%%%%%%%%%%%%%%%%%%%

\section{Introduction}

In this article we try to give a condensed panoramic view
of the development of two-di\-men\-si\-o\-nal rational conformal field theory in
the last twenty-five years. Given its limited length, our contribution
can be, at best, a collection of pointers to the literature.
Needless to say, the exposition is highly biased, by the taste and
the limited knowledge of the authors. We present in advance our apologies to
everyone whose work has been inappropriately represented or even omitted.

The study of conformal field theories in two dimensions has its origin in
several distinct areas of physics:

\def\leftmargini{1.1em}~\\[-2.4em] \begin{itemize}\addtolength\itemsep{-3pt}

\item 
In the attempt to describe the strong interactions in elementary particle 
physics in the framework of dual models.

\item 
Under the label string theory, these models were reinterpreted as a perturbation 
series containing a gravitational sector. Conformal field theories appear as 
theories defined on the world sheet that is swept out by the string \cite{diSc}.

\item 
In statistical mechanics, conformal field theory plays a fundamental role
in the theory of two-di\-men\-si\-o\-nal critical systems \cite{poly5,frqs2}
by describing fixed points of the renormalization group.

\item 
Conformal field theories on surfaces with boundary
arise in quasi one-di\-men\-si\-o\-nal condensed matter physics 
in the description of impurities, e.g.\ in the Kondo effect \cite{affl6}.

\item 
Chiral conformal field theories describe universality classes of quantum Hall 
fluids \cite{fcgkkmst}; conformal blocks are used as approximate wave functions 
for quasi-particles \cite{moRe}.

\end{itemize}

In the last 25 years, two-di\-men\-si\-o\-nal rational conformal field 
theory has been moreover a major area of cross-fertilization between 
mathematics and physics. Many fields of mathematics have both benefitted
and contributed to conformal field theory. Examples include representation 
theory, infinite-di\-men\-si\-o\-nal algebra, the theory of modular forms, 
algebraic and differential geometry, and algebraic topology.

\medskip

We find it helpful to divide the era we are discussing into three periods. The 
first is what we will refer to as the classical period, comprising roughly the 
second half of the 1980s, in which much fundamental insight into conformal field 
theories was obtained. This was followed by a period of subsequent consolidation, 
which essentially comprises the first half of the nineties. The third period 
started basically with the advent of D-branes in string theory, which had a 
significant impact on the field: it gave a strong boost to the study of boundary 
conditions, which resulted in a much deeper understanding of the structure of 
rational conformal field theory. This period of research has brought (higher) 
category theoretic structures to the forefront, both in algebraic and in 
geometric approaches to rational conformal field theory.
We end with a discussion of some topics of current interest
and with a list of omissions.
  
There is a close relation between conformal field theory in two
dimensions and topological field theory in three dimensions. 
In fact, both for chiral conformal field theory and for full local conformal 
field theory, many structural aspects become much more transparent by using 
insights obtained from three-di\-men\-si\-o\-nal topology. For this 
reason, some of the developments in three-di\-men\-si\-o\-nal
topological field theory are covered in this review as well.

%%%%%%%%%%%%%%%%%%%%%%%%%%%%%%%%%%%%%%%%%%%%%%%%%%%%%%%%%%%%%%%%%%%%%%%%

\section{Groundbreaking insights}\label{s1}

Historically, the classical period has its roots in the theory of
dual resonance models in the early seventies 
and in the study of operator product expansions in quantum field theory. 
Important pieces of insight gained already back then are the following:

\def\leftmargini{1.1em}~\\[-2.4em] \begin{itemize}\addtolength\itemsep{-3pt}

\item 
Chiral quantities -- in particular conserved chiral currents -- 
live on a double cover $\widehat X$ of the surface $X$ on
which the conformal field theory is considered \cite{ales}.
The surface $\widehat X$ is naturally oriented and inherits a conformal
structure from $X$. In two dimensions, these data are equivalent to
a {\em holomorphic\/} structure on $\widehat X$. This relationship is at the
basis of powerful complex-analytic techniques available for the study
of two-di\-men\-si\-o\-nal conformal field theories.

\item 
The requirement of conformal invariance yields strong constraints on 
the structure of quantum field theory. For instance, when combined with the 
principle of locality, it leads to the L\"uscher-Mack theorem \cite{luma2,Fust} 
which gives the most general commutation relations among the components of 
the stress-energy tensor. The modes of the stress-energy tensor are generators 
of an infinite-di\-men\-si\-o\-nal Lie algebra, the Virasoro algebra. This
forms the basis for the use of techniques from infinite-di\-men\-si\-o\-nal
algebra in conformal field theory.

\item 
Operator product expansions can be studied with the help of
representation theoretic methods see e.g.\ \cite{DMppt,ruYu}.
Indeed, conformal field theory in general has been one of the
driving forces for the representation theory of infinite-di\-men\-si\-o\-nal
Lie algebras and related structures.

\item 
Current symmetries induce conformal symmetries, via a variant
of the Sugawara construction \cite{baha,dafr,somM'}.
  
\item
Low-dimensional field theories allow for `exotic' statistical behavior of quantum 
fields more general than the one of bosons or fermions \cite{stWil,lemy}, now
referred to as anyon statistics.
Also, bosonic theories can have sectors which behave like fermions, and vice versa 
\cite{cole5,froh3,fren2}; this is known as the boson-fermion correspondence.

\end{itemize}

\noindent
The 1980s have witnessed impressive progress in two-di\-men\-si\-o\-nal 
conformal field theories. To give an idea of these developments, we
refer to classics in the literature and, where appropriate,
indicate on which previous developments the respective article is
based. A list of highlights is, in chronological order:

\def\leftmargini{1.1em}~\\[-2.4em] \begin{itemize}\addtolength\itemsep{-3pt}

\item 
In 1984, Belavin, Polyakov and Zamolodchikov \cite{bepz} used 
infinite-di\-men\-si\-o\-nal symmetries to reduce the description of operator
algebra expansions to a finite-di\-men\-si\-o\-nal problem and computed the 
operator algebra expansions for bulk fields in Virasoro minimal models. Chiral 
symmetry structures that are strong enough to allow for such a reduction from 
infinite- to finite-di\-men\-si\-o\-nal problems are one of the hallmarks of 
{\em rational\/} conformal field theories.

The work of \cite{bepz} strongly relies on results from the study
of dual models, in particular about the structure of the conformal
group and of the Virasoro algebra. One crucial input is information about
highest weight representations of the Virasoro algebra from the
work of Feigin and Fuks \cite{fefu} and Kac \cite{kac17}. 

\item 
Also in 1984, Witten \cite{witt20} introduced a Lagrangian formulation of
conformal field theories with current algebra symmetries, now known as 
Wess-Zu\-mino-Wit\-ten (WZW) models. This takes the form of a sigma model with 
target space a group manifold, as in the previously studied principal chiral model 
(see e.g.\ \cite{deeM,powi}), but includes 
a Wess-Zumino term, i.e.\ \cite{wezu,novi} a multi-valued topological action. 

\item 
In the same year, Knizhnik and Zamolodchikov \cite{knza} derived differential 
equations on the correlators of the same class of models by studying 
combined null vectors of the affine Lie algebra and the Virasoro algebra.

\item 
In 1986, Goddard, Kent and Olive used the coset construction \cite{goko2} to 
realize the chiral unitary Virasoro minimal models. The coset construction is, 
even to date, a major source for rational chiral conformal field theories. 
In addition, it
supplies examples of quantum field theories with gauge symmetries that 
are at the same time exactly tractable using methods from representation theory.

\item 
Also in 1986, Borcherds \cite{borC3} introduced the notion of a vertex algebra 
to mathematics as a formalization of the physical concept of a chiral algebra,
i.e.\ the algebraic structure encoding the chiral symmetries  of a conformal 
field theory. His motivation was to gain a better understanding of ``monstrous 
moonshine'', a conjecture connecting the largest sporadic group, the monster 
group, to modular forms. A vertex algebra can be thought of as a generalization 
of a $\mathbb{Z}_+$-graded commutative associative unital algebra equipped with 
a derivation 
of degree 1 with a multiplication depending on a formal coordinate.
Various concepts from conformal field theory and string theory are
instrumental for Borcherds' work. In particular, the no-ghost theorem
of Goddard and Thorn \cite{goth} ensures that the Lie algebras associated
to the vertex algebras of interest are indeed generalized Kac-Moody algebras.
The physical principle of locality appears in the definition of vertex algebras 
as a requirement on the commutator of chiral fields. 

\item 
And again in 1986, Cardy \cite{card3} showed that imposing modular invariance 
of the torus partition function constitutes a powerful constraint on the field 
content of a conformal field theory. One year later, Cappelli, Itzykson and Zuber 
\cite{caiz} presented a classification of modular invariant partition functions 
for the $\mathfrak{su}(2)$ WZW model and for Virasoro minimal models. 
The modular properties of characters of affine Lie algebras computed by Kac
and Peterson \cite{kape3} are crucial input.
These classifications display an A-D-E pattern 
similar to the classification of simply laced complex simple Lie algebras,
of discrete subgroups of $SU(2)$, or of elementary catastrophes.

\item 
In 1987 Fr\"ohlich \cite{Froh,froh11} emphasized the relevance of the
braid group for the statistical properties of quantum fields in two and three
dimensions, and their relations with monodromy properties of correlation functions
and with the Yang-Baxter equation. Closely related observations about statistical 
properties and the exchange relations for chiral fields were made by Rehren and 
Schroer \cite{resC}, Tsuchiya and Kanie \cite{tska2}, and Longo \cite{long3}. 
Braid group statistics includes anyon statistics as a special case, namely 
the one in which the braid group representation is one-dimensional.

\item 
In 1988-1989 Moore and Seiberg \cite{mose,mose3,mose4} unraveled much of the 
categorical structure underlying chiral rational conformal field theories. 
Their work was based on the results about braid group statistics just
mentioned and on the idea of Friedan-Shenker \cite{frsh2} to formulate 
two-dimensional conformal field theory as analytic geometry on a universal 
moduli space of Riemann surfaces. It extended results about the fusion rules
that had been obtained by E.\ Verlinde \cite{verl2} using heuristic arguments 
about the factorization of conformal blocks on elliptic curves. Felder, 
Fr\"ohlich and Keller \cite{fefk2} introduced a family of coproducts on chiral 
algebras to gain a representation theoretic understanding of these findings.

\item 
Also in 1989, Fredenhagen, Rehren and Schroer \cite{frrs} discussed the 
superselection structure of sectors with braid group statistics in
low-di\-men\-si\-o\-nal quantum field theory.
The aspects particular two two-di\-men\-si\-o\-nal \emph{conformal}
field theory were analyzed in \cite{frrs2,gaFr}. These studies are 
performed in the algebraic framework for quantum field theory, 
which was developed by Doplicher, Haag and Roberts \cite{dohr,dohr3} and others 
and is, in turn, based on earlier work of Haag and Kastler \cite{haka}.
A crucial role in the algebraic framework is played by the locality principle,
which provides a conceptually clear understanding of the appearance of
braid group statistics in low dimensional quantum field theory.

\item 
At the same time, Witten \cite{witt50} developed non-abelian Chern-Simons 
theory, a three-di\-men\-si\-o\-nal topological field theory which yields 
invariants for links that can be specialized to the Jones polynomial.
The Jones polynomial can equally be recovered from differential equations on 
expectation values of Wilson lines that can be derived in a functional integral 
approach, see \cite{frki}. Witten's results generalize in particular earlier 
work by  A.S.\ Schwarz \cite{schw'} on abelian gauge theories having a 
Chern-Simons term only. Witten also shows how structures of chiral 
rational conformal field theory, e.g.\ the Verlinde formula for the fusion 
rules, becomes transparent when viewed from three-di\-men\-si\-o\-nal geometry.

The Chern-Simons three-form is closely connected to characteristic classes. 
When Chern-Simons theory is evaluated on special three-manifolds, it provides
invariants of knots and links. Thereby Chern-Simons theory provides several
links between conformal field theory and algebraic topology.

At about the same time, Atiyah presented an axiomatization of 
topological quantum field theory \cite{atiy6} that is based on
Segal's proposal \cite{sega9} for an axiomatics of conformal field theory. 

\item 
Again in 1989, Cardy \cite{card9} exhibited a close relation between the 
Verlinde formula and boundary conditions of a rational conformal field theory
with modular invariant partition function given by charge conjugation.
\end{itemize}

\noindent
Numerous reviews and books provide details about the work in the classical 
period. An (again biased) selection is \cite{Fust,Algs,Mose,DIms,Sche2',gago}.

%%%%%%%%%%%%%%%%%%%%%%%%%%%%%%%%%%%%%%%%%%%%%%%%%%%%%%%%%%%%%%%%%%%%%%%%

\section{Extension and consolidation}

The papers listed in section \ref{s1} continue to stimulate much of the 
ongoing research in the field, both in mathematics and in physics. Here we 
present a very brief summary of some important subsequent developments.

Subsection \ref{ss31} gives an an overview of chiral symmetry structures.
An important aspect of the representation theory of these structures is 
that they allow the construction of vector bundles with connection on the 
moduli space of complex curves, called conformal blocks; aspects of this theory
are discussed in subsection \ref{ss32}. The system of vector bundles formed
by the conformal blocks is essentially the mathematical formalization
of chiral conformal field theory. This is a highly non-trivial system of 
bundles; it allows one to endow the representation category of 
the chiral symmetry algebra with additional structure.
In the strongest possible situation, one obtains the structure of a modular 
tensor category, which is the subject of subsection \ref{ss33}.

Full local conformal field theories -- which arise e.g.\ as the quantization of
certain Lagrangian field theories, or in the continuum limit of certain lattice models
-- can be constructed from chiral conformal
field theories. A specific ingredient of this construction is to find a modular
invariant torus partition function. This is the subject of subsection \ref{ss34}.
Lagrangian approaches both to conformal field theory and to topological field
theory are very briefly discussed in subsections \ref{ss35} and \ref{ss36}, 
respectively.

%%%%%%%%%%%%%%%%%%%%%%%%%%%%%%%%%%%%%%%%%%%%%%%%%%%%%%%%%%%%%%%%%%%%%%%%

\subsection{Chiral symmetry structures}\label{ss31}

As a first important example of a vertex algebra, Frenkel, Lepowsky and Meurman 
\cite{FRlm} constructed the monster vertex algebra, whose group of vertex algebra 
automorphisms is the monster sporadic group. This vertex algebra plays a crucial 
role in Borcherds' proof of the moonshine conjecture (for reviews, see e.g.\ 
\cite{GAnn,godd5}). The monster vertex algebra is actually a conformal vertex 
algebra, i.e.\ it comes with a Virasoro element, which gives rise to a field whose 
modes furnish a representation of the Virasoro algebra on any module over the 
vertex algebra.

Much activity was aimed at constructing further classes of vertex 
algebras or chiral symmetry algebras in different formalizations. 
Some major lines of development were the following.

\def\leftmargini{1.1em}~\\[-2.4em] \begin{itemize}\addtolength\itemsep{-3pt}

\item
A strong driving force was the quest for larger symmetry structures which, while 
having Virasoro central charge larger than one, still yield rational theories. 
One of the resulting notions is the structure of a $\cal W$-algebra; for a review 
of such algebras we refer to \cite{Bosc}.  

\item
By general arguments, every modular invariant partition function can be constructed 
in two steps \cite{mose4}: in the first step the chiral algebras for left- and 
right-moving degrees of freedom are extended, yielding chiral theories with 
isomorphic fusion rules. The pairing of left- and right-movers in the local 
theory then requires one to select an isomorphism, and this choice constitutes the
second step. Extensions of chiral symmetry algebras are therefore of much 
interest. Two classes of extensions are particularly well understood:
conformal embeddings \cite{babo,scwa} of chiral algebras, which are generated by
non-abelian currents, and extensions by representations of quantum dimension one,
so-called simple currents \cite{scya6,bant7,dolm}. 

\item
The original construction of the monster vertex algebra involved the extension 
of an orbifold of some lattice vertex algebra involving the Leech lattice. The 
orbifold construction \cite{dhvw} continues to be a rich source of algebraic
\cite{dvvv} and geometric \cite{hava} constructions. It turns out that the 
relevant category of world sheets for an orbifold theory with orbifold group 
$G$ are $G$-coverings, which can have branch points at the insertions of
twist fields. The product of $N$ copies of a vertex algebra carries an obvious
action of any subgroup $G\,{\subseteq}\,\mathfrak S_N$. The corresponding orbifold 
theory is called a permutation orbifold. Permutation orbifolds enter 
crucially in a proof of the congruence subgroup conjecture \cite{bant14}.

\item
The coset construction \cite{goko2} provides a particularly important 
generalization of vertex algebras based on affine Lie algebras. It also admits
a Lagrangian description \cite{gaku3} and continues to be an important testing 
ground for new concepts in conformal field theory. For most coset 
conformal field theories the problem of selection rules and field 
identification, and of resolving fixed points under this identification,
arises. In spite of progress on the representation theory of the coset chiral 
algebra, both in terms of the underlying affine Lie algebras \cite{fusS5} 
and concerning category theoretic aspects \cite{ffrs2}, this problem is not 
completely solved. Irrational conformal field theories \cite{hkoc} provide a 
hint to a large generalization of the coset construction; for the time being,
these theories offer, however, more challenges than results.

\end{itemize}

On the other hand, the classification of 
vertex algebras, or just of rational conformal vertex algebras, is not a realistic 
goal. For instance, every even lattice, in particular every even self-dual one,
provides such a vertex algebra. Even self-dual lattices only exist in 
dimensions $d$ which are a multiple of $8$. It follows from the Siegel-Minkowski 
mass formula that already in dimension 32 there are at least $10^7$ such 
lattices. The corresponding conformal vertex algebras have a single irreducible
representation and Virasoro central charge equal to $d$. More generally,
conformal vertex algebras with a single irreducible representation must
have a Virasoro central charge $c$ which is a multiple of $8$. For
$c\,{=}\,24$ so-called meromorphic conformal field theories, which are supposed 
to correspond to such vertex algebras, were classified in \cite{sche5}.

\medskip

The vertex algebras relevant for \emph{rational} conformal field theory have a 
semisimple representation category. In this case various aspects of the representation 
category are well understood, including in particular the Huang-Lepowsky theory 
of tensor products for categories of modules over a conformal vertex algebra
(for a review see \cite{hule3.5}). Much of the work of the Huang-Lepowsky
school is built on this theory. A related approach to tensor products is given 
by the Gaberdiel-Kausch-Nahm algorithm \cite{gaKa,nahm8}. An extension of tensor 
product theory to the non-semisimple case was developed in \cite{hulz}.

For several reasons, tensor categories associated with affine Lie algebras are 
of particular interest: they are related to problems in algebraic geometry, and 
they have strong connections with representation categories of quantum groups.
Properties of these categories were implicitly obtained already in 1986 through 
the study of null vector equations \cite{knza} for the corresponding conformal 
blocks. In the early nineties, the tensor structure for theories with negative
level was constructed by Kazhdan and Lusztig \cite{kaLuX}; this result was 
transferred to positive level by Finkelberg \cite{fink}. It can also be 
recovered in the more general Huang-Lepowsky theory of tensor products.

\medskip

Vertex algebras have meanwhile also become an important tool in several areas of 
pure mathematics, including the algebraic geometry of Hilbert schemes \cite{leso} 
and the geometric Langlands program (see \cite{freN12} for a review). 
Sheaves of vertex algebras also enter in the chiral de Rham complex
\cite{masv}, an attempt for a geometric realization of elliptic cohomology.
The locality principle that is integrated in the definition of vertex algebras 
has also provided a fruitful rationale to single out 
structures in a wider context of infinite-dimensional algebra \cite{kac38}.

%%%%%%%%%%%%%%%%%%%%%%%%%%%%%%%%%%%%%%%%%%%%%%%%%%%%%%%%%%%%%%%%%%%%%%%%

\subsection{Conformal blocks}\label{ss32}

An important aspect of vertex algebras is that they give rise to {\em conformal 
blocks\/}, i.e.\ vector bundles over moduli spaces of curves with marked points. 
If the vertex algebra is a conformal vertex algebra, then these bundles carry a 
projectively flat connection, called Knizhnik-Zamolodchikov connection.
For vertex algebras based on affine Lie algebras, these vector bundles 
were first explored by Tsuchiya, Ueno and Yamada \cite{tsuY,ueno8}. 
In this case they provide interesting non-abelian generalizations of
theta functions and are thus of independent interest in algebraic geometry, 
see e.g.\ the Bourbaki talk \cite{sorg} and \cite{beau8}.
 %NB: CFT bourbakabel - e.g.  \freN5, \gawe2, \jone11, \math7, \sorg
 %                      sowie \bost, \voge2, freN20, sega7

The ranks of these bundles are given by the Verlinde formula; their computation 
is highly non-trivial. Various different arguments, including complete proofs,
have been given for the Verlinde formula for Wess-Zumino-Witten models:

\def\leftmargini{1.1em}~\\[-2.4em] \begin{itemize}\addtolength\itemsep{-3pt}
\item 
Heuristic arguments, based on a path integral approach to Chern-Simons theories
\cite{dawe,blth}.
\item 
Techniques from algebraic geometry \cite{falt3,beau8}. (For earlier work 
restricted to $\mathfrak{sl}(2)$, see also \cite{besz,thad2}.)
\item 
Fixed points of loop group actions \cite{almw1}.
\item 
Techniques from homological algebra, in particular Lie algebra cohomology, combined
with a vanishing argument \cite{tele}.
\item 
Holomorphic quantization of Chern-Simons theories \cite{axdw,jewe} and related
methods from symplectic geometry \cite{mewo}.
\end{itemize}

For the construction of conformal blocks from vertex algebras we refer to the
Bourbaki talk \cite{freN5} and to the textbook \cite{FRbe}. This construction is 
also crucial for the relation to the Beilinson-Drinfeld theory of chiral algebras
\cite{BEdr}. Conversely, from the values of all $n$-point conformal blocks on a 
suitable finite-di\-men\-si\-o\-nal subspace of states, a vertex algebra can be
reconstructed \cite{gago}.

\medskip

Since one deals with bundles of conformal blocks, and thus multivalued functions 
instead of single-valued correlators, this theory is not an ordinary physical 
quantum field theory. Still, it appears directly in applications in physics, such 
as in the description of universality classes of quantum Hall fluids (for a review 
see \cite{fpsw}). It should be appreciated that the direction of the magnetic 
field distinguishes a chirality in a quantum Hall fluid; accordingly the relevance 
of chiral (as opposed to full local) conformal field theory is rather natural.

\medskip

\noindent
There are two major approaches to make conformal blocks explicitly computable.

\def\leftmargini{1.1em}~\\[-2.4em] \begin{itemize}\addtolength\itemsep{-3pt}

\item 
Based on a construction of modules over the Virasoro algebra by Feigin and Fuks 
\cite{fefu6}, Dotsenko and Fateev \cite{doFa} were able to calculate the 
fusing matrices and bulk structure constants in Virasoro minimal models. The 
underlying idea \cite{feld} is to construct modules over complicated
symmetry structures as the BRST cohomology of free field representations.
This leads in particular to explicit expressions for solutions of the
Knizhnik-Zamolodchikov equations, see e.g.\ \cite{feVa}. The BRST approach 
also brought up several Kazhdan-Lusztig type correspondences \cite{fesT,fgst4}.

\item 
Another approach uses differential equations, which can be obtained e.g.\ from 
null vectors in representations of the chiral algebra. Examples include the 
Knizhnik-Zamolodchikov equation \cite{knza} (for reviews, see \cite{ETfk,math7})
and the Gep\-ner-Wit\-ten equation \cite{gewi}. Zhu introduced \cite{zhu3} a 
finiteness condition on vertex algebras, called $C_2$-cofiniteness, which does 
not only guarantee the convergence of the characters of modules over a vertex algebra 
\cite{zhu3}, but also leads to good differential equations for chiral blocks. The 
same cofiniteness condition implies the existence of Zhu's algebra \cite{frzh} which 
constitutes one important tool for the study of chiral conformal field theory.

The differential equations for conformal blocks can be made most explicit for 
surfaces of low genus. For results in genus one see e.g.\ \cite{bern2}, and for 
the particular case of differential equations for characters see e.g.\ \cite{mams3}.

\end{itemize}

%%%%%%%%%%%%%%%%%%%%%%%%%%%%%%%%%%%%%%%%%%%%%%%%%%%%%%%%%%%%%%%%%%%%%%%%

\subsection{Modular tensor categories}\label{ss33}

The categorical structure found in the work of Moore and Seiberg leads to the
notion of a modular tensor category. A rigorous construction of
three-di\-men\-si\-o\-nal topological field theory or, equivalently, of a modular 
functor, by Reshetikhin and Turaev \cite{retu,retu2} is based on this notion.

Modular tensor categories also arise naturally in the approach to conformal 
field theory via the Doplicher-Haag-Roberts framework. Indeed, the category of 
local sectors of a net of von Neumann algebras on the real line of finite 
$\mu$-index is a (unitary) modular tensor category if the net is 
strongly additive and has the split property \cite{kalm}.

In the vertex algebra approach to conformal field theory, the relevant result is
that the representation category of a self-dual vertex algebra that obeys the
$C_2$-cofiniteness condition and certain conditions on its homogeneous subspaces 
is a modular tensor category, provided that this category is semisimple 
\cite{huan21}. The result of \cite{huan21} relies on a careful implementation of 
the structures unraveled by Moore and Seiberg, combined with Huang-Lepowsky's 
theory of tensor products of representations of conformal vertex algebras.

The structural importance of the $C_2$-cofiniteness condition cannot be 
overrated. Recent studies of vertex algebras with non-semisimple representation 
category (see e.g.\ \cite{miya8}) indicate that $C_2$-cofiniteness suffices, even
in the absence of semisimplicity, to endow the representation category
with a structure reasonably close to the one of a modular tensor category.
This opens the way to a better understanding of logarithmic \cite{gura}, and 
possibly also other non-rational, conformal field theories.

\medskip

Let us make two more comments on the relation between modular
tensor categories and vertex algebras:

\def\leftmargini{1.1em}~\\[-2.4em] \begin{itemize}\addtolength\itemsep{-3pt}

\item 
It is worth emphasizing that for endowing the representation category of a 
suitable conformal vertex algebra with the structure of a modular tensor 
category, only properties of conformal blocks for surfaces of genus $0$ and $1$
are used. Now from conformal vertex algebras and modular tensor categories one 
obtains representations of mapping class groups in two different ways. First,
via the Knizhnik-Zamolodchikov connection on the bundles of conformal blocks. 
In this case the mapping class group arises as the fundamental group of
the moduli space. And second, via three-di\-men\-si\-o\-nal topological 
field theory and the embedding of mapping class groups in cobordisms. The 
latter construction uses directly the modular tensor category, while the former 
is based on the vertex algebra and its conformal blocks. To establish that, 
for any genus, the two constructions give equivalent representations will 
require a serious improvement of the understanding of conformal blocks for 
rational vertex algebras.

\item 
For lattice vertex algebras, the modular tensor category together with the
value of the Virasoro central charge is equivalent to the genus of the lattice.
The genus of a lattice $L$ is, by definition, the collection of all local lattices
$L\,{\otimes_{\mathbb Z}}\,{\mathbb Z}_p$, including $L\,{\otimes_{\mathbb Z}}\,
\mathbb R$. For a Euclidean lattice $L$, the latter is equivalent to the quadratic 
space $({L^*}\!{/}\!{L},q)$ given by the discriminant form and the rank of the 
lattice. In conformal field theory terms these data correspond to the equivalence
class, as a braided monoidal category, of the representation category of the
lattice vertex algebra and to the value of the Virasoro central charge.
\\
This suggests \cite{hohn6} to regard the modular tensor category and the Virasoro 
central charge of a given rational vertex algebra as arithmetic information and 
raises, in particular, the question of whether good Mass formulas exist for 
vertex algebras.
\end{itemize}

The categorical dimensions of simple objects in a modular tensor category 
are typically not integers. As a consequence, modular tensor categories do not 
admit fiber functors to the category of complex vector spaces. However, one can 
show that fiber functors to categories of bimodules over a suitable ring exist.
As that ring, one can take the endomorphism ring of any generator of a module 
category over the modular tensor category. Reconstruction then yields algebraic 
objects like weak Hopf algebras \cite{bosz} or other ``quantum symmetries'' 
\cite{masc3,ocne10}. Conversely, it has been established \cite{nitv} that connected 
ribbon factorizable weak Hopf algebras over $\mathbbm C$ with a Haar integral have 
representation categories which are modular tensor categories.

The tensor product functor on a modular tensor category $\cal C$ is exact. As 
a consequence, the Grothendieck group $K_0(\cal C)$ carries a natural structure
of a ring, the so-called fusion ring. It has a natural basis given by the 
classes of simple objects of $\cal C$. Verlinde's formula \cite{verl2} (which 
for general rational conformal field theories was proven in \cite{huan21})
relates the structure constants of the fusion ring in this basis to the modular
transformations of the characters of vertex algebra modules. This relationship 
motivated the study of various aspects of fusion rings, which was actively 
pursued in the early 1990s; for a review see \cite{jf24}.

%%%%%%%%%%%%%%%%%%%%%%%%%%%%%%%%%%%%%%%%%%%%%%%%%%%%%%%%%%%%%%%%%%%%%%%%

\subsection{Classification of modular invariant partition functions}\label{ss34}

Intrigued by the A-D-E-structure that was found in the classification of modular
invariants for minimal models and $\mathfrak{su}(2)$ WZW models, in the early 
nineties several groups pursued the program to classify modular invariant torus 
partition functions. On the one hand, the theory of simple currents, due to 
Schellekens and Yankielowicz, gave a powerful machinery to construct modular 
invariant partition functions. (See \cite{scya6} for a review, and \cite{krSc} 
for the general classification.) At that point, exceptional modular invariants, 
i.e.\  modular invariants that are not explainable via simple currents and charge 
conjugation, remained still rather mysterious. On the other hand, Gannon and 
others were able to obtain remarkable classifications for some particular 
classes of models; see e.g.\ \cite{gann3,stan,gann5,garw2,cogR}.
   
This classification program is put into a different 
perspective by the observation (see e.g.\ \cite{scya5,vers,fusS})
   % in \scya5 eq. (B7), in \vers appendix B, in \fusS sec 4. cp also \gann17
that there exist modular invariant bilinear combinations of characters (with 
non-negative integral coefficients and with unique vacuum) 
which are unphysical in the sense that they cannot arise as 
the torus partition function of any consistent local conformal field theory: 
Finding all such bilinear combinations of characters provides very useful 
restrictions on the possible form of torus partition functions, but is not 
equivalent to classifying consistent conformal field theories.
(The existence of spurious solutions should not come as a surprise, since 
many more conditions are to be satisfied: the sewing constraints, which 
implement compatibility of the correlators under cutting and gluing of surfaces,
see e.g.\ \cite{sono3,lewe3,prss3}.) Strikingly,
the true diagonal modular invariant is not necessarily physical (for a counter 
example see \cite{sosc}); a classification of those modular tensor categories 
for which the diagonal modular invariant is physical is still unknown.

%%%%%%%%%%%%%%%%%%%%%%%%%%%%%%%%%%%%%%%%%%%%%%%%%%%%%%%%%%%%%%%%%%%%%%%%

\subsection{Sigma model approaches to conformal field theory}\label{ss35}

The motivation to consider sigma model approaches, i.e.\ models based on 
spaces of maps $\Sigma \,{\to}\, M$, include extending WZW models 
to noncompact groups \cite{gawe6} as well as understanding cosmologically 
interesting backgrounds in string theory \cite{divv6}. Sigma models have provided 
strong links between quantum field theory and various aspects of geometry, in 
particular complex and symplectic geometry, leading \cite{hitc16} to common 
generalizations of both.

There is a vast literature on conformal sigma models, which we are unable to 
review appropriately. We restrict our attention to developments that have direct 
relation to \emph{rational} conformal field theory. 
Early important work by Felder, Gaw\c edzki and Kupiainen \cite{fegk} obtains
in a path integral approach the quantization of the level
and the bulk spectra of Wess-Zumino-Witten sigma models on
compact groups that are not necessarily simply connected.
For an extension to coset conformal field theories, we refer to \cite{gaku3,kaSc2}.

String vacua based on rational conformal field theories were first constructed 
by Gepner \cite{gepn5}; his construction was later generalized by Kazama and 
Suzuki \cite{kasu2}. Remarkably, many of these models can be matched with string 
compactifications 
on Calabi-Yau manifolds. The symmetry induced by conjugation 
of the $U(1)$ charge in a rational $N\,{=}\,2$ superconformal field theory is in 
this way at the basis of mirror symmetry of Calabi-Yau spaces \cite{levw}.

%%%%%%%%%%%%%%%%%%%%%%%%%%%%%%%%%%%%%%%%%%%%%%%%%%%%%%%%%%%%%%%%%%%%%%%%

\subsection{Topological field theory from path integrals and state sums}\label{ss36}

Witten's work on Chern-Simons theories has been extended in various ways. Because 
of the close relation between chiral conformal field theory and 
three-di\-men\-si\-o\-nal topological field theory, we mention here some aspects of 
this development. Topological field theories based on finite groups were 
considered by Dijkgraaf and Witten \cite{diwi2}; the paper presents structural 
relations to WZW theories based on non-simply connected compact Lie groups.

Lagrangian descriptions of topological field theories have been influential in 
several directions. For instance, Chern-Simons perturbation theory gave rise to 
Vassiliev invariants (for a review see \cite{voge2}). The theory of Vassiliev 
invariants has found applications in other fields of mathematics as well, e.g.\ 
to universal Lie algebras \cite{voge3} and to holomorphically symplectic 
manifolds in the form of Rozanski-Witten \cite{rowi} invariants. 
  
Another construction of topological field theories is via state sum models.
These have been discussed in various dimensions, including lattice topological 
field theories in two \cite{fuhk} and three \cite{chfs} dimensions, as well as 
the Turaev-Viro construction \cite{tuvi}. 
The latter implements an old idea of Ponzano and Regge \cite{pore} to build 
invariants with the help of $6j$-symbols for tensor categories and yields, in 
the case of modular tensor categories, an invariant of three-manifolds that 
is the absolute square of the Reshetikhin-Turaev invariant.

%%%%%%%%%%%%%%%%%%%%%%%%%%%%%%%%%%%%%%%%%%%%%%%%%%%%%%%%%%%%%%%%%%%%%%%%

\section{New frontiers and categorical structures}

Dual models started out as theories describing hadrons as bound states of charged 
particles which were supposed to be located at the end points of an open string. 
As a consequence, the underlying surfaces on which the theory is considered were
allowed to have non-empty boundary. In contrast, the first two periods of the
development of rational conformal field theory covered in this review were 
largely concerned only with closed orientable surfaces. But in order to study
defects in systems of condensed matter physics \cite{osaf}, percolation 
probabilities \cite{card12}, (open) string perturbation theory in the background 
of the string solitons known as D-branes \cite{polc8}, and order-disorder 
dualities, it became again necessary to consider also surfaces that may have 
boundaries and\,/\,or can be non-orientable, as well as surfaces with 
defect lines.

More specifically, the advent of D-branes in string theory brought boundary 
conditions in rational conformal field theories to the center of interest. Based
on earlier work of Cardy \cite{card9} and of the Rome group
(in particular \cite{prss3}, for a review see \cite{Ansa}), it became
evident that boundary conditions provide crucial new
{\em structural\/} insight into conformal field theories. In particular,
the fundamental importance of the two-fold oriented cover $\widehat X$ of $X$
(which had already been noted in the study of dual models, see section
\ref{s1} above) gained again fundamental importance:
full local conformal field theory on a conformal
surface $X$ is related to a chiral conformal field theory on $\widehat X$.

%%%%%%%%%%%%%%%%%%%%%%%%%%%%%%%%%%%%%%%%%%%%%%%%%%%%%%%%%%%%%%%%%%%%%%%%

\subsection{Structure constants}\label{ss:sc}

Early progress in the study of conformal field theory on surfaces with boundary
focussed on the computation of structure constants for operator products.
Similarly as in the case of closed surfaces, for which structure
constants had been obtained (see e.g.\ \cite{doFa,zafa2,chfl,fukl,pezu})
by analyzing four-point correlators of bulk fields on the sphere,
the first complete results for the structure constants on surfaces with 
boundaries were obtained for Virasoro minimal models \cite{runk,runk2}.
   
A strong focus was on the computation of one-point correlators of bulk fields 
on a disk, which may be collected in the coefficients of
so-called boundary states \cite{ishi}. They 
provide significant information of much interest for applications, like ground 
state degeneracies \cite{aflu} or Ramond-Ramond charges of string compactifications 
\cite{bDlr}. Moreover, they encode the integral coefficients of the annulus 
partition functions and thus give the spectrum of boundary fields. Ramond-Ramond 
charges are related to twisted K-theory \cite{mimo,witt112}. This leads to the 
interpretation of the Verlinde algebra of Wess-Zumino-Witten models as equivariant 
twisted K-theory of compact, connected and simply-connected Lie groups (see 
\cite{free12} for a review), and to conformal field theory techniques for 
computing D-brane charges \cite{frsc,brSch2}.

In string compactifications based on a rational conformal field theories, such 
as those obtained by Gepner's \cite{gepn5} construction, the boundary states of 
the conformal field theory correspond to specific D-branes in the string theory 
(see e.g.\ \cite{bDlr}). The construction of boundary states for such theories 
turns out to be remarkably subtle. In the case of Gepner models, it was first 
discussed in \cite{reSC}. For the complete construction, including a correct 
treatment of field identification fixed points, see \cite{fusw} and \cite{fkllsw}.

An important structural observation \cite{bppz} concerns
the annulus coefficients: they furnish matrix-valued representations
with non-negative integer entries (so called NIMreps) of the fusion ring.
Even more important was the insight \cite{bppz2,fffs} that the constraints on 
structure constants for boundary fields preserving a given boundary condition 
give rise to generalized pentagon relations. The mixed $6j$-symbols appearing 
in these relations have the conceptual interpretation as multiplication morphisms 
of an associative algebra in the modular tensor category of chiral data.
This algebra has the physical interpretation as the algebra of boundary
fields for a fixed boundary condition.

%%%%%%%%%%%%%%%%%%%%%%%%%%%%%%%%%%%%%%%%%%%%%%%%%%%%%%%%%%%%%%%%%%%%%%%%

\subsection{TFT, Frobenius algebras and CFT correlators}\label{ss:TFT-CFT}

In the rational case the algebra of boundary fields turns out to have the 
mathematical structure of a special symmetric Frobenius algebra in the modular 
tensor category of chiral data. This Frobenius algebra is the crucial ingredient 
in a construction that describes the correlation functions of rational conformal 
field theories (RCFTs) in terms of invariants provided by three-dimensional 
topological field theories (TFTs); for a review see \cite{scfr2}. Special symmetric 
Frobenius algebras also appear in the approach to conformal field theory via nets 
of von Neumann algebras, where they are realized \cite{lore,evpi,lore2} as 
so-called $Q$-sys\-tems \cite{long6}.
    
In the description via TFT, the RCFT correlators are realized as invariants of 
ribbon graphs in three-manifolds under the Reshetikhin-Turaev modular functor. 
Thereby geometric structure in three dimensions is again used to gain insight into
two-di\-men\-si\-o\-nal theories, in this case into full {\em local\/} conformal 
field theories. The relevant three-manifold geometry also appears in the study of 
CFT on world sheet orbifolds with the help of Chern-Simons theory \cite{hora7}.

The description of full local CFT via TFT allows one in particular to recover 
modular invariant partition functions. The problem (which has spurious unphysical
solutions) of classifying bilinear combinations of characters as candidates for 
modular invariant partition functions is thereby replaced \cite{fuRs4} by the 
problem of classifying Morita classes of special symmetric Frobenius algebras or,
equivalently \cite{ostr}, appropriate module categories. 
Moreover, the formalism treats exceptional modular 
invariants and simple current invariants on the same footing. For instance, 
for the tensor categories based on the $\mathfrak{sl}(2,\mathbbm C)$ Lie algebra, 
the Cappelli-Itzykson-Zuber A-D-E classification is recovered \cite{kios}. 

The bulk theory is obtained from the Frobenius algebra of the boundary theory
as a commutative special symmetric Frobenius algebra in the product of the modular 
tensor category of the RCFT with its opposed category. This relation between 
boundary and bulk generalizes the situation found \cite{moSe} in 
two-di\-men\-si\-o\-nal topological field theories, where both the boundary 
Frobenius algebra and the bulk Frobenius algebra are algebras in the category 
of complex vector spaces. This structure can also be interpreted in terms of 
vertex algebras, leading to so-called full field algebras \cite{huKo2}
and open-closed field algebras \cite{kong6}.

The TFT construction provides a detailed dictionary between algebraic notions
and physical concepts. A generalization of an algebra with involution, a so-called 
Jandl structure on $A$, allows one to construct also correlators for unoriented 
surfaces. Semisimple Frobenius algebras with an involution had appeared before in 
the study of two-di\-men\-si\-o\-nal lattice topological field theories 
\cite{kamo}. Symmetries and order-disorder dualities of RCFT correlators can be 
conveniently described with the help of defect lines \cite{ffrs5}, which in turn 
have a natural representation within the TFT formulation of the CFT correlators.

%%%%%%%%%%%%%%%%%%%%%%%%%%%%%%%%%%%%%%%%%%%%%%%%%%%%%%%%%%%%%%%%%%%%%%%%

\subsection{Some current lines of research}

In the last couple of years, new and important lines
of research in conformal field theory have come to the foreground.
Here, we give a choice that is, of course, again biased.

\subsubsection{Beyond semisimple rational conformal field theories}

At present, many efforts are devoted to a better understanding of
conformal field theories based on chiral symmetry structures whose representation
categories are not semisimple any longer. 

The most important example is logarithmic conformal field theory, a generalization 
of rational conformal field theory dating back to the early nineties \cite{gura}. 
Much insight has been gained for the particular class of $(1,p)$-models for which 
a Kazhdan-Lusztig correspondence to a Hopf algebra exists (see \cite{gaTi} for 
a recent summary). Logarithmic conformal field theories are also important 
for their close relation to models of statistical mechanics, e.g.\ to 
percolation models \cite{resa3}.    

Other important subclasses of theories are non-compact theories like Liouville 
theory and its relatives (for a review, see \cite{tesc9}) and sigma models with 
supersymmetric target spaces \cite{sasc} which currently find applications both in 
the context of the AdS/CFT duality \cite{abjm} and in statistical mechanics, e.g.\ 
in the description of transitions between quantum Hall plateaux \cite{zirn6}.

%%%%%%%%%%%%%%%%%%%%%%%%%%%%%%%%%%%%%%%%%%%%%%%%%%%%%%%%%%%%%%%%%%%%%%%%

\subsubsection{Higher categorical geometry for target spaces}

In Lagrangian approaches to conformal field theories, higher categorical 
structures have become more and more important as well. Already in the mid 
eighties, Alvarez \cite{alvaO5} and Gaw\c edzki \cite{gawe1} noticed that the 
Wess-Zumino term is closely related to Deligne hypercohomology. Brylinski 
\cite{BRyl}, Murray \cite{murr} and others then constructed hermitian bundle 
gerbes with connection as a geometric realization of hypercohomology. This 
allows in particular for a geometric understanding of the Wess-Zumino term 
as a surface holonomy for these gerbes. This description was extended to D-branes 
by the notion of gerbe modules \cite{gawe18} including non-abelian effects, to
holonomy for unoriented surfaces with the introduction of Jandl structures for 
gerbes \cite{scsW}, and to a target space interpretation of
defect lines with the notion of bibranes \cite{fusW}.

One current point of interest are extensions to topological field theories that 
also associate values to 1- and 2-manifolds. They require higher category theory, 
see e.g.\ \cite{fhlt} for a recent discussion based on compact Lie groups. An 
early contribution to this area was published in this journal \cite{baDo}.
Higher categories and higher-di\-men\-si\-o\-nal algebra have become
an increasingly important topic of research, including
the idea \cite{crfr} to use categorification to extend techniques
from three-di\-men\-si\-o\-nal topological field theory to higher dimensions.

Another area in which category theory and conformal field theory meet is the 
study of D-brane categories. To review their role in homological mirror 
symmetry \cite{kont7} is beyond the scope of our survey; for an exposition that 
emphasizes conformal field theory aspects in terms of sigma models, we refer 
to \cite{aspi11}.

The relation between three-di\-men\-si\-o\-nal topological and
two-di\-men\-si\-o\-nal conformal field theory leads to a corresponding 
relation between higher categorical structures: the hermitian bundle gerbe 
with connection on a compact connected group $G$ describing the Wess-Zumino term 
of the two-di\-men\-si\-o\-nal theory is a 2-categorical structure. If this gerbe 
admits a multiplicative structure, it can be lifted \cite{cjmsw} to a 2-gerbe on 
the classifying space of $G$, and thus to a 3-categorical structure. The 
three-holonomy of a corresponding 2-gerbe with connection describes the Chern-Simons 
term of the associated three-di\-men\-si\-o\-nal topological field theory.

%%%%%%%%%%%%%%%%%%%%%%%%%%%%%%%%%%%%%%%%%%%%%%%%%%%%%%%%%%%%%%%%%%%%%%%%

\subsubsection{SLE}

Another field of mathematics that shed a new light on conformal field theory 
are stochastic differential equations. Stochastic Loewner Evolution (SLE) 
was defined by Schramm \cite{schra3}; for reviews see e.g.\ \cite{wernW,card26}.
SLE (or, more precisely, chordal SLE) provides a measure on curves on the upper 
half plane which start at zero and end at infinity, as well as on domains conformally 
equivalent to the upper half plane. The measure satisfies three conditions: conformal 
invariance, reflection symmetry, and a restriction property. These properties turn 
out to fix the measure up to a real parameter $\kappa\,{\ge}\,0$, which is related 
to the Virasoro central charge by $c\,{=}\,(3\kappa{-}8)(6{-}\kappa){/}(2\kappa)$.

In certain examples it has been established that the SLE measure on curves 
arises as a continuum limit of the weight of spin configurations with a prescribed 
domain boundary in two-dimensional lattice models.
One such example is critical percolation, where crossing 
formulas conjectured by Cardy \cite{card12} and by Watts \cite{watt6} using boundary 
conformal field theory  were proved in \cite{smirS3} and \cite{dube3}, respectively.

The factorization of amplitudes is expected from the heuristic path integral 
formulation of CFT. The approaches to CFT reported in sections \ref{ss:sc} and 
\ref{ss:TFT-CFT} provide a mathematical formalization of this property. Approaches based 
on SLE, on the other hand, directly provide a microscopic description by replacing 
the path integral by a precisely defined measure. This provides a much more 
direct relation to lattice models, 
but it is by no means straightforward to express CFT correlators in the SLE formulation. 
For example, conjecturally, correlators with insertions of the stress-energy tensor 
can be obtained from the probability that the SLE curve passes 
through a number of small intervals centered at the insertion points \cite{doRc} 
(this is at central charge $c\,{=}\,0$; for $c$ between 0 and 1 see \cite{doyo2}),
and there is a relation between Virasoro null vectors and SLE martingales \cite{baBer5}.

%%%%%%%%%%%%%%%%%%%%%%%%%%%%%%%%%%%%%%%%%%%%%%%%%%%%%%%%%%%%%%%%%%%%%%%%

\section{Omissions}

There are important aspects of two-dimensional rational conformal field 
theories that we could not cover in this review.
Some of them are the following:

\begin{itemize}

\item 
Applications of conformal field theory to entanglement entropy \cite{caCa6}.

\item 
Applications of conformal field theory to quantum computing \cite{nssfd}. 

\item 
D-branes in WZW models, in particular their target space
geometry, \cite{alsc2,fffs}.

\item 
Renormalization group flows between conformal field theories, including
Zamolodchikov's $c$-theorem \cite{zamo7} and an analogous result for flows 
between conformal boundary conditions \cite{aflu, frko}. 

\item 
Integrable \cite{zamO5,ghza} and numerical \cite{yuZa,dPtw} methods for 
investigating bulk and boundary perturbations of conformal field theories.

\item
The generalization of the fermion-boson correspondence to higher genus world
sheets (see e.g.\ \cite{amnvb} for an early discussion), and applications
of the fermion-boson correspondence in combinatorics, see e.g. \cite{beMS}.

\item 
Galois symmetries acting on chiral data \cite{deGo}, and their
use in the study of modular invariants \cite{coga,fusS} 
and in the proof of the congruence subgroup conjecture \cite{bant14}. 

\item
The relation between fusion rings and quantum cohomology, see e.g. \cite{gepn11,witt87}.

\item
Applications of conformal and superconformal field theory
in elliptic cohomology \cite{sega7,stte}.

\end{itemize}

\noindent
That the topics in this list, and more, have not even been treated briefly in this
survey illustrates what an enormous amount of knowledge about rational conformal 
field theory has been compiled during the last quarter century.

%%%%%%%%%%%%%%%%%%%%%%%%%%%%%%%%%%%%%%%%%%%%%%%%%%%%%%%%%%%%%%%%%%%%%%%%

%\vskip 4.5em
 \newpage

\noindent{\sc Acknowledgments:} \\
We would like to thank J.\ Fr\"ohlich and M.\ Gaberdiel for helpful comments.
\\
JF is supported partially by VR under project no.\ 621-2006-3343.
CS is supported partially by the Collaborative Research Center 676 ``Particles,
Strings and the Early Universe - the Structure of Matter and Space-Time''.

%%%%%%%%%%%%%%%%%%%%%%%%%%%%%%%%%%%%%%%%%%%%%%%%%%%%%%%%%%%%%%%%%%%%%%%%

%\newpage
 \vskip 4.5em

\raggedright

 \newcommand\wb{\,\linebreak[0]}
 \def\wB {$\,$\wb}
 \newcommand\Bi[1]    {\bibitem{#1}}
 \newcommand\JO[6]    {{\em #6}, {#1} {#2}, {#4}-{#5} ({#3}) }
 \newcommand\JP[4]    {{\em #4}, \href{http://#3}{#1} {#2} }
 \newcommand\JX[7]    {{\em #7}, \href{http://dx.doi.org/#6}{#1} {#2}, {#4}-{#5} ({#3}) }
 \newcommand\JY[7]    {{\em #7}, \href{http://#6}{#1} {#2}, {#4}-{#5} ({#3}) }
 \newcommand\KK[7]    {{\em #7}, {#1} {#2}, {#4}-{#5} ({#3})
                      \href{http://arXiv.org/abs/#6} {\tt[#6]} }
 \newcommand\KU[8]    {{\em #8},\\ \href{http://dx.doi.org/#6}{#1} {#2}, {#4}-{#5} ({#3})
                      \href{http://arXiv.org/abs/#7} {\tt[#7]} }
 \newcommand\KX[8]    {{\em #8}, \href{http://dx.doi.org/#6}{#1} {#2}, {#4}-{#5} ({#3})
                      \href{http://arXiv.org/abs/#7} {\tt[#7]} }
 \newcommand\KY[8]    {{\em #8}, \href{http://#6}{#1} {#2}, {#4}-{#5} ({#3})
                      \href{http://arXiv.org/abs/#7} {\tt[#7]} }
 \newcommand\KM[9]    {{\em #9}, \href{http://dx.doi.org/#6}{#1} {#2}, {#4}-{#5} ({#3})
                      \href{http://arXiv.org/abs/#8} {\tt[#8\,(#7)]} }
 \newcommand\Prep[2]  {{\em #2}, preprint \href{http://arXiv.org/abs/#1} {\tt #1} }
 \newcommand\Preq[3]  {{\em #3}, preprint \href{http://arXiv.org/abs/#2} {\tt #2\,(#1)} }
 \newcommand\Pret[3]  {{\em #3}, preprint \href{http://#2} {\tt #1} }
 \newcommand\Preu[2]  {{\em #2}, preprint {\tt #1} }
 \newcommand\BOOK[4]  {{\em #1\/} ({#2}, {#3} {#4})}
 \newcommand\BOOX[5]  {\href{http://#2}{\em #1\/} ({#3}, {#4} {#5})}
 \newcommand\inBO[8]  {{\em #8}, in:\ {\em #1}, {#2}\ ({#3}, {#4} {#5}), p.\ {#6}-{#7} }
 \newcommand\inBK[9]  {{\em #9}, in:\ \href{http://#7}{\em #1}, {#2}\ ({#3}, {#4} {#5}), p.\ {#6} 
                      \href{http://arXiv.org/abs/#8} {\tt[#8]}}
 \newcommand\inBL[9]  {{\em #9}, in:\ {\em #1}, {#2}\ ({#3}, {#4} {#5}), p.\ {#6}-{#7}
                      \href{http://arXiv.org/abs/#8} {\tt[#8]} }
 \newcommand\inBX[9]  {{\em #9}, in:\ \href{http://#8}{\em #1}, {#2}\ ({#3}, {#4} {#5}), p.\ {#6}-{#7}}
 \newcommand\inBY[9]  {{\em #9}, in:\\ \href{http://#8}{\em #1}, {#2}\ ({#3}, {#4} {#5}), p.\ {#6}-{#7}}
 \def\adma  {Adv.\wb in Math.}
 \def\anip  {Ann.\wb Inst.\wB Poin\-car\'e}
 \def\anop  {Ann.\wb Phys.}
 \def\aspm  {Adv.\wb Stu\-dies\wB in\wB Pure\wB Math.}
 \def\aste  {Ast\'e\-ris\-que}
 \def\atmp  {Adv.\wb Theor.\wb Math.\wb Phys.}
 \def\bams  {Bull.\wb Amer.\wb Math.\wb Soc.}  
 \def\clqg  {Class.\wb Quant.\wb Grav.}
 \def\comp  {Com\-mun.\wb Math.\wb Phys.}
 \def\crap  {C.\wb R.\wb Acad.\wb Sci.\wB Paris (S\'erie I -- Math\'ematique)}
 \def\duke  {Duke\wB Math.\wb J.}
 \def\fiic  {Fields\wB Institute\wB Commun.}
 \def\foph  {Fortschr.\wb Phys.} 
 \def\focm  {Found.\wb Comput.\wb Math.}
 \def\fuaa  {Funct.\wb Anal.\wb Appl.}
 \def\gafa  {Geom.\wB and\wB Funct.\wb Anal.}
 \def\icmb  {Doc.\wb Math.\wb J.\wb DMV\wb Extra\wb Volume\wb ICM}
 \def\ihes  {Publ.\wb Math.\wB I.H.E.S.}
 \def\ijmp  {Int.\wb J.\wb Mod.\wb Phys.\ A}
 \def\ijmb  {Int.\wb J.\wb Mod.\wb Phys.\ B}
 \def\imrn  {Int.\wb Math.\wb Res.\wb Notices}
 \def\inma  {Invent.\wb math.}
 \def\injm  {Int.\wb J.\wb Math.}
 \def\isjm  {Israel\wB J.\wb Math.}
 \def\jams  {J.\wb Amer.\wb Math.\wb Soc.}  
 \def\jetl  {Sov.\wb Phys.\wB JETP\wB Lett.}
 \def\jgap  {J.\wb Geom.\wB and\wB Phys.}
 \def\jhep  {J.\wb High\wB Energy\wB Phys.}
 \def\jlms  {J.\wB London\wB Math.\wb Soc.}
 \def\joag  {J.\wB Al\-ge\-bra\-ic\wB Geom.} 
 \def\jofa  {J.\wb Funct.\wb Anal.}
 \def\jodg  {J.\wb Diff.\wb Geom.}
 \def\jomp  {J.\wb Math.\wb Phys.}
 \def\jopa  {J.\wb Phys.\ A}
 \def\josp  {J.\wb Stat.\wb Phys.}
 \def\jsyg  {J.\wB Sym\-plec\-tic\wB Geom.}
 \def\lemp  {Lett.\wb Math.\wb Phys.}  
 \def\mpla  {Mod.\wb Phys.\wb Lett.\ A}
 \def\nupb  {Nucl.\wb Phys.\ B}
 \def\nuci  {Nuovo\wB Cim.}
 \def\phlb  {Phys.\wb Lett.\ B}
 \def\phrd  {Phys.\wb Rev.\ D}
 \def\phrl  {Phys.\wb Rev.\wb Lett.}
 \def\phrp  {Phys.\wb Rep.}
 \def\phrv  {Phys.\wb Rev.}
 \def\pnas  {Proc.\wb Natl.\wb Acad.\wb Sci.\wb USA}
 \def\ptrf  {Probab.\wB Theory\wB Related\wB Fields}
 \def\qjmo  {Quart.\wb J.\wb Math.\wB Oxford}
 \def\remp  {Rev.\wb Mod.\wb Phys.}     
 \def\rinc  {Riv.\wB Nuovo\wB Cim.}
 \def\rvmp  {Rev.\wb Math.\wb Phys.}
 \def\sebo  {S\'emi\-nai\-re\wB Bour\-baki}
 \def\sema  {Selecta\wB Mathematica}
 \def\sjnp  {Sov.\wb J.\wb Nucl.\wb Phys.}
 \def\slnm  {Sprin\-ger\wB Lecture\wB Notes\wB in\wB Mathematics}
 \def\slnp  {Sprin\-ger\wB Lecture\wB Notes\wB in\wB Physics}
 \def\somd  {Sov.\wb Math.\wb Dokl.}
 \def\taia  {Top\-o\-lo\-gy \wB and\wB its\wB Appl.}
 \def\topo  {Topology}
 \def\trgr  {Trans\-form.\wB Groups}
 \def\tujm  {Tur\-kish\wB J.\wb Math.}

\end{document}